\documentclass[fleqn,twoside,twocolumn,nofootinbib,showkeys]{revtex4}
 
\usepackage[sec,nocpr]{ujp} 

\begin{document}
\title[Vortex structures and electron beam dynamics in magnetized plasma]
{VORTEX STRUCTURES AND ELECTRON BEAM DYNAMICS IN MAGNETIZED PLASMA}%

\author{V.I.~Maslov}
\affiliation{Space Research Institute,  Nat. Acad. of Sci. of Ukraine  
 and the State Space Agency of Ukraine}
\address{40, 4/1, Glushkov Ave  Kyiv 187, 03680, Ukraine}
\affiliation{NSC Kharkov Institute of Physics and Technology, Nat. Acad. of Sci. of Ukraine}
\address{1, Academichna Str.,  Kharkiv 61108, Ukraine}
\email{vmaslov@kipt.kharkov.ua}
\author{O.K.~Cheremnykh}
\affiliation{Space Research Institute,  Nat. Acad. of Sci. of Ukraine  
 and the State Space Agency of Ukraine}
\address{40, 4/1, Glushkov Ave  Kyiv 187, 03680, Ukraine}
\author{A.P.~Fomina}
\affiliation{Bogolyubov Institute for Theoretical Physics, Nat. Acad. of Sci. of Ukraine}%
\address{14b, Metrolohichna Str., Kyiv 03143, Ukraine}%
\author{R.I.~Kholodov}
\affiliation{Institute of Applied Physics,  Nat. Acad. of Sci. of Ukraine}%
\address{58, Petropavlivska Str., 40000 Sumy, Ukraine}%
\author{O.P.~Novak}
\affiliation{Institute of Applied Physics,  Nat. Acad. of Sci. of Ukraine}%
\address{58, Petropavlivska Str., 40000 Sumy, Ukraine}%
\author{R.T.~Ovsiannikov}
\affiliation{Karazin Kharkiv National University}%
\address{4, Svobody Sq., 61022 Kharkiv, Ukraine}%

\udk{533} \pacs{52.35.We,  94.20.wf} \razd{}

\autorcol{V.I.~Maslov, O.K.~Cheremnykh, A.P.~Fomina et al.}

\setcounter{page}{1}%

\begin{abstract}

The formation of vortex structures at reflection of electron beam from the double layer of the Jupiter ionosphere is investigated in this paper. And also the influence of these vortex structures on the formation of dense upward electron fluxes, accelerated by the double layer potential along the Io flux tube is studied. Then a phase transition to the cyclotron superradiance mode becomes possible for these electrons. The conditions of the vortex perturbations formation are considered. The nonlinear equation is found that describes the vortex dynamics of electrons and its consequences are studied.

\end{abstract}


\maketitle

\section{Introduction}

In this paper the vortex dynamics of electrons are considered and the processes that can influence on formation of upward electron fluxes in plasma double layers (DL) are studied in the framework of the cyclotron superradiance (CSR) effect. The possibility of vortex structures formation due to the interaction of electron currents with the Jupiter ionosphere plasma is discussed.

The influence of Io on Jupiter`s  magnetosphere and its radio emission has been studied for a long time \cite{1,2}. This active satellite is constantly erupting particles into the ionosphere and is connected with the generation of strong radio emission from Jupiter, discovered in 1955. The interaction between Io and Jupiter occurs in the form of electric currents that moves from Io along the Jupiter magnetic field lines and close through the ionosphere near the planet poles, forming Io magnetic flux tube (IFT). Near the footprint of the IFT, where the particles carrying this current influence on the Jupiter atmosphere, auroras arise. They are observed in the form of bright tails of radiation stretched behind the main spot along the movement of the IFT footprint  on the planet surface \cite{3,4}. Besides the projections of other satellites are also visible, but of much lower brightness. The radiation spot of Io at a certain point has the shape of ellipse with a 	size $\sim$200-500 km, which is slightly larger than the projection size of the satellite itself onto the  planet near $\sim$200 km (due to the IFT constriction). It indicates that the Io-Jupiter interaction region is slightly larger than the size of Io. IFT tail is less bright in the northern hemisphere, where the magnetic field is stronger than in the southern hemisphere. And conversely, the radio emission is more intense in northern areas \cite{5,6}. Jupiter's equatorial field strength is 4.3 gauss, and ranging from 10 gauss at the south pole to 14 gauss  at the north pole. 

Since 2016 NASA's  Juno spacecraft has been orbiting Jupiter, equipped with devices to compile detailed information about the planet. Based on these data, new effects were discovered, for example, the observations report of distinct, high-energy, discrete electron acceleration in Jupiter's auroral polar regions and also about upward magnetic-field-aligned electric potentials of up to 400 keV, an order of magnitude larger than the largest potentials observed at Earth.  Also Juno's Energetic particle Detector Instrument (JEDI) detected intense electron beams moving away from Jupiter's polar regions. In these beams there are often found  electrons with energies above $\sim$1 MeV, sometimes up to $>$10 MeV. These beams occur primarily above the swirl region of the polar cap aurora. It have found a correlation between the swirl emergence from Ultraviolet Spectrograph (UVS) and the very intense beams from JEDI \cite{4,7,8}. While in literature additional mysteries still exist on the precise nature of this acceleration. 

In works \cite{10,11,12,13,14} the original model was proposed that describes the generation of a super-powerful Jovian radio emission, which is based on the effect of the Fomin--Dicke collective coherent CSR for a system of inverted electrons at high Landau levels in rarefied magnetized plasma. According to the model, electron beams are accelerated from Io toward Jupiter and are reflected from double electric layers that arise in the plasma of the ionosphere. Then the reflected electron beams move upward and pass into the coherent superradiance CSR mode. The criterion of the phase transition to the CSR mode depends, in particular, on such characteristics of electron beams as density and temperature.  

The DL phenomenon has been studied for a long time (see \cite{15,16,17,18,19,20,21}) and observed in laboratory conditions and  in astrophysical applications.  Also, DL are observed in auroral regions where it require some external driver to produce electron acceleration \cite{22,23}. Jupiter's auroras also, according to observations, sometimes have parallel potential DL, analogous to Earth's aurora. The electric DL are plasma regions with a violated quasineutrality and with  size of several Debye radii. The formation of such layers is possible at the border between regions in plasma with different characteristics, for example, with different temperatures. The formation mechanisms of sufficiently strong DL are also possible, associated with the injection of particles into the plasma, due to the current amplification of the plasma instability. Most of the beam electrons in the ionosphere are reflected by  parallel electrical layer and cause auroras, although some of the electrons are scattered.

Numerical simulation in \cite{14} showed the possibility of a quasi-stationary DL formation during the interaction of an electron beam with a plasma in the nonrelativistic case. When  electron beam with a density of $10^{4}$cm$^{-3}$ is injected into the plasma of the Jupiter ionosphere, a DL can form in the region where densities of the beam and plasma are approximately equal. The phase portrait of DL and the distribution of free and trapped components of the electron beam were found. It also shows the formation of cold electron beams, for which a transition to the CSR mode is possible. A characteristic potential drop is observed in the DL region. The DL potential is shown in Fig.1.

\begin{figure}
\includegraphics[width=\column]{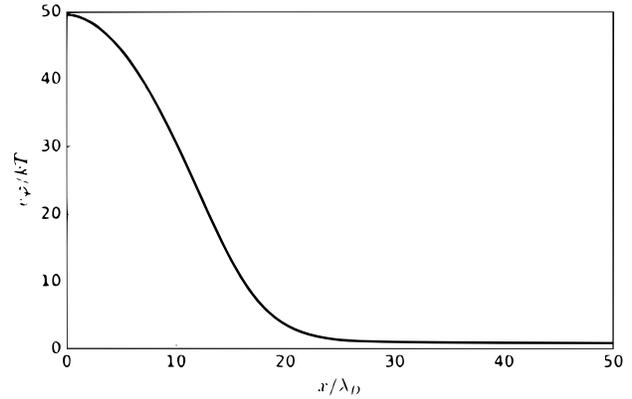}
\vskip-3mm
\caption{Potential profile of the double layer}
\end{figure}

The presented distribution was obtained by averaging the values of  potential and field strength over a time interval that exceeding much the period of plasma oscillations. The instantaneous values can be significantly distorted due to plasma oscillations. As seen in Fig.1 the DL has width of approximately 20~$r_{de}$. The potential drop in the DL is determined by the energy of the electron beam. For example, for the following parameters of plasma electrons - concentration n$_{0} ~\approx  10^{4}$~cm$^{-3}$ and temperature kT~$\approx  1 $~keV, we get the Debye radius $r_{de} \approx 2.4$~m, plasma frequency $\omega_{p}^{-1}\approx 1.8~\cdot 10^{-7}$~s and, accordingly, the width of  DL is $\sim$50~ m, the time of formation $\sim 10^{-6}$~s. In this case the energy of the beam is 50~keV, which corresponds to the potential of DL  50~kV. 

Numerical simulation of double layer was carried out in the region of 100~$r_{de}$ with open boundary conditions. At the initial moment of time it was assumed that the region is filled with an equilibrium plasma with a temperature kT and an electron concentration $n_{0}$, the ion component of the plasma is frozen and uniformly distributed. A continuous electron beam was injected into the plasma from the left at a rate equal to $10v_{T}$. Within a short time, approximately equal to $30/ \omega_{p}$, a quasistationary mode is established in the phase space of the system with DL reflecting a part of the beam back and small oscillations in the plasma. A typical view of  phase portrait is shown in Fig.~2. 

\begin{figure}
\includegraphics[width=\column]{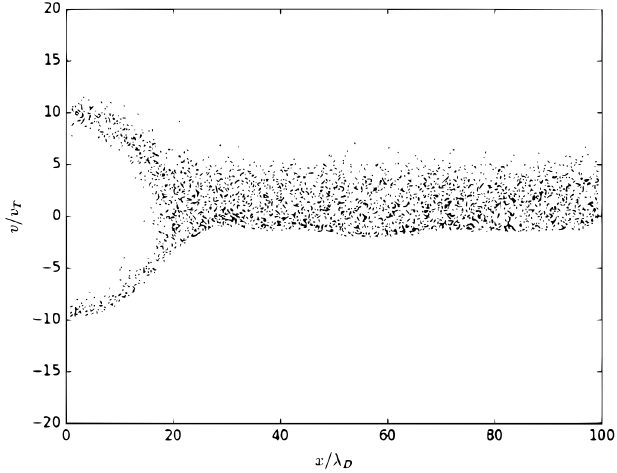}
\vskip-3mm\caption{The phase portrait of the double layer in plasma}
\end{figure}

In \cite{24,25} it was shown that when an electron beam from Io enters Jupiter's ionosphere, beam-plasma instability (BPI) develops. The electron distribution function becomes wider due to the excited fields and then these electrons cause the ultraviolet aurora. Since BPI develops locally in the inhomogeneous plasma, it can lead to the formation of DL at a certain height. The properties and stability of the formed intense DL, as well as the dynamics of plasma particles, are described. It is noted that the reflection of electron beams from the Jupiter atmosphere can lead to the formation of a semi-vortex. The effect of the space charge of a decelerated reflected beam and its collision with particles of partially ionized plasma lead to a gradual expansion of the beam. Thus the beam is reflected back with a large radius, which can lead to vortex formation.

\section{
 Vortex in a plane (r, z) in a collisional magnetized plasma with a reflected electron beam}

\noindent

Consider the feasible vortex dynamics of a reflected electron beam near Jupiter in a plane, one of the axes of which is directed along  the magnetic field under the presence of electron collisions.

We assume that axis z is directed towards the surface of Jupiter. Since the electric double layer has a finite longitudinal size along z, approximately equal to $L_{DL} \approx {\sqrt{2} V_{b} / \omega_{pe} } $ 
\cite{15,16,17,19}. And also, due to the finite radius of the beam, the DL has a finite transverse dimension. Then the DL has not only a longitudinal electric field $E_{z}$, but also a transverse electric field $E_{r}$.

Although  electrons are magnetized, the beam expands in $E_{r}$  due to collisions and forms an obvious half-vortex in the process of reflection from the DL, under  the condition when the radius of a half-vortex exceeds the radius of the beam, that falls on the layer. We will estimate this condition. Since the crossed configuration of the radial electric $E_{or} $ and longitudinal magnetic $H_{o} $ fields is maintained in the vicinity of the DL, then the electron beam, due to collisions of electrons with a frequency $\nu_{e}$, expands with a velocity
\begin{equation} 
\label{EQ1} 
V_{r} =-\frac{e\nu _{e} E_{or} }{m_{e} \omega _{He}^{2} } ,        
\end{equation} 
where  field $E_{or} $ is determined by the space charge of the DL, 
$\omega _{He} =eH_{o} / m_{e} c $ is the cyclotron frequency of electrons. For close longitudinal and radial dimensions of the double layer, we have approximately $E_{or} \approx E_{oz} $. Since the amplitude of the electric potential of the DL is approximately equal to the kinetic energy of the beam, and the length  is approximately equal to 
$L_{DL} \approx {\sqrt{2} V_{b} / \omega_{pe} } $ 
\cite{15,16,17,19}, 
then for the formation of a half-vortex, one can obtain the condition
\begin{equation} 
\label{EQ2} 
\frac{\nu _{e} }{\omega _{He}^{2} } \frac{\omega _{pe} }{\pi } \approx 1.         
\end{equation} 
So, 
$\omega _{pe} ~\equiv ~ \left( 4\pi n_{oe} e^{2} /   m_{e}  \right)^{1/2} $ 
- the electron plasma frequency 	must be greater than $\omega _{He} $, if $\omega _{He} $ is greater than $\nu _{e} $.

\section{
 Vortex in the plane (r, $\theta$) in the collisionless approximation in a magnetized plasma with a reflected electron beam}
 
\noindent 

Now we consider the vortex dynamics of a reflected electron beam near Jupiter in a plane orthogonal to the magnetic field. Since in the vicinity of DL there is a crossed configuration of the radial electric $E_{or} $and longitudinal magnetic field $H_{o} $, then vortices can also form in the plane (r, $\theta$), because the nonequilibrium  state is maintained due to the drift of electrons along the angle $\theta $ with a velocity
\[
V_{\theta o} =-\frac{eE_{or} }{m_{e} \omega _{He} } =\left(\frac{\omega _{pe}^{2} }{2\omega _{He} } \right)\left(\frac{\Delta n}{n_{oe} } \right)r\equiv r\omega _{\theta o} .\] 
\[\Delta n\equiv n_{oe} -q_{i} {n_{oi}  \mathord{\left/{\vphantom{n_{oi}  e}}\right.\kern-\nulldelimiterspace} e}.
\] 
where $q_{i} $, $n_{oi} $ - ion charge and density.

Consider a vorticity $\alpha $ - an electron vortex characteristic
\begin{equation} \label{EQ3} 
\alpha \equiv \vec{e}_{z} rot\vec{V}=\frac{1}{r} \partial _{r} rV_{\theta } -\frac{1}{r} \partial _{\theta } V_{r}.     
\end{equation} 
The physical sense of   $\alpha $becomes obvious if we introduce the angular velocity of electron rotation  in a vortex $\Omega \equiv V_{\theta} /r $ , then 
\begin{equation} \label{EQ4} 
\alpha =2\Omega +r\partial _{r} \Omega -\frac{1}{r} \partial _{\theta } V_{r} ~.       
\end{equation} 
If $\Omega \ne \Omega (r) $ и $V_{r} =0$, then the vorticity is equal to the double angular velocity of electron rotation $\alpha~=~2\Omega $.

\section{
Equations describing the excitation of nonlinear vortex perturbations}

Let us obtain the equation that describes the excitation and properties of vortex perturbations. 
In \cite{26} derivation of the equation for the vorticity begins with the momentum equation (2) for a viscous, Newtonian conducting fluid, in which there is no explicitly electric field. We use the hydrodynamic equations for electrons at times shorter than the capture with taking into account electron collisions
 \[
 \frac{\partial \vec{V}}{\partial t} +\nu _{e} \vec{V}+\left(\vec{V}\vec{\nabla }\right)\vec{V}= 
\]
\begin{equation}\label{EQ5}
=\left(\frac{e}{m_{e} } \right)\vec{\nabla }\varphi +\left[\vec{\omega }_{He} ,\; \vec{V}\right]-\left(\frac{V_{th}^{2} }{n_{e} } \right)\vec{\nabla }n_{e},  
\end{equation}
\[
\frac{\partial n_{e} }{\partial t} +\vec{\nabla }\left(n_{e} \vec{V}\right)=0 
\]
and the Poisson equation for the electric potential $\varphi $
\begin{equation} 
\label{EQ6}  
\Delta \varphi =4\pi \left(en_{e} -q_{i} n_{i} \right),       
\end{equation} 
here $\vec{V}$, $n_{e} $ - electron velocity and density, $V_{th} $ - thermal electron velocity, $\vec{V}_{i}$, $n_{i}$, $q_{i}$ - ion velocity, density and charge. As we will see below the dimensions of vortex disturbances are much larger than the Debye electron radius $r_{de} \equiv V_{th} / \omega_{pe}$, then we can neglect the last term in (\ref{EQ5}). 

We obtain a unified nonlinear equation describing the vortex dynamics of electrons. For this we use rot for (\ref{EQ5}), i.e. we act vectorially by the operator $\vec{\nabla }\times $ on (\ref{EQ5}). Then we get
\begin{equation} 
\label{EQ7} 
\frac{\partial \vec{\alpha }}{\partial t} +\nu _{e} \vec{\alpha }+\left[\vec{\nabla }\times \left(\vec{V}\vec{\nabla }\right)\vec{V}\right]=\left[\vec{\nabla }\times \left[\vec{\omega }_{He} \times \vec{V}\right]\right] 
\end{equation} 
Here $\vec{\alpha }=\left[\vec{\nabla }\times \vec{V}\right]$. To transform the last equation we use the expression
\[
\left[\vec{\nabla }\times \left[\vec{\omega }_{He} \times \vec{V}\right]\right]=\left(\vec{\nabla }\vec{V}\right)\vec{\omega }_{He} -\left(\vec{\nabla }\vec{\omega }_{He} \right)\vec{V}= 
\]
\begin{equation} 
\label{EQ8} 
=\vec{\omega }_{He} \left(\vec{\nabla }\vec{V}\right)+\left(\vec{V}\vec{\nabla }\right)\vec{\omega }_{He} -\left(\vec{\omega }_{He} \vec{\nabla }\right)\vec{V} 
\end{equation} 
at $\vec{\nabla }\vec{\omega}_{He}=0$ and 
\begin{equation} 
\label{EQ9} 
\left[\vec{V}\times \left[\vec{\nabla }\times \vec{V}\right]\right]=\left[\vec{V}\times \vec{\alpha }\right]=1/2 \vec{\nabla }V^{2} -\left(\vec{V}\vec{\nabla }\right)\vec{V}.
\end{equation} 
From (\ref{EQ9}) we get the expression
\[
\left[\vec{\nabla }\times \left(\vec{V}\vec{\nabla }\right)\vec{V}\right]=-\left[\vec{\nabla }\times \left[\vec{V}\times \vec{\alpha }\right]\right]= 
\]
\[
=-\left(\vec{\nabla }\vec{\alpha }\right)\vec{V}+
\left(\vec{\nabla }\vec{V}\right)\vec{\alpha }= 
\]
\begin{equation} 
\label{EQ10} 
=\left(\vec{V}\vec{\nabla }\right)\vec{\alpha }+\vec{\alpha }\left(\vec{\nabla }\vec{V}\right)-\left(\vec{\alpha }\vec{\nabla }\right)\vec{V}-\vec{V}\left(\vec{\nabla }\vec{\alpha }\right) 
\end{equation} 
From (\ref{EQ7}), (\ref{EQ8}), (\ref{EQ10}) and  $\vec{\nabla }\vec{\alpha }=0$ we find
\[
\partial _{t} \vec{\alpha }+\nu _{e} \vec{\alpha }+\left(\vec{V}\vec{\nabla }\right)\vec{\alpha }+\vec{\alpha }\left(\vec{\nabla }\vec{V}\right)-\left(\vec{\alpha }\vec{\nabla }\right)\vec{V}=  
\]
\begin{equation} 
\label{EQ11} 
=\vec{\omega }_{He} \left(\vec{\nabla }\vec{V}\right)+\left(\vec{V}\vec{\nabla }\right)\vec{\omega }_{He} -\left(\vec{\omega }_{He} \vec{\nabla }\right)\vec{V} 
\end{equation} 
Hence we have
\[
d_{t} \left(\vec{\alpha }-\vec{\omega }_{He} \right)+\nu _{e} \vec{\alpha }+\left(\vec{\alpha }-\vec{\omega }_{He} \right)\left(\vec{\nabla }\vec{V}\right)= 
\]
\begin{equation} 
\label{EQ12} 
=\left(\left(\vec{\alpha }-\vec{\omega }_{He} \right)\vec{\nabla }\right)\vec{V} 
\end{equation} 
\[
d_{t} \equiv \partial _{t} +\left(\vec{V}\vec{\nabla }\right).
\] 
We transform the third term of the left side of (\ref{EQ12}) as follows 
\[
\left(\vec{\alpha }-\vec{\omega }_{He} \right)\left(\vec{\nabla }\vec{V}\right)=-\left(\frac{\vec{\alpha }-\vec{\omega }_{He} }{n_{e} } \right) \cdot \nonumber \\ 
\]
\begin{equation} 
\label{EQ13}
 \cdot \left[\partial _{t} +\left(\vec{V}\vec{\nabla }\right)\right]n_{e} =-\left(\frac{\vec{\alpha }-\vec{\omega }_{He} }{n_{e} } \right)d_{t} n_{e}  
\end{equation} 
From  (\ref{EQ12}), (\ref{EQ13}) we find
\begin{equation} 
\label{EQ14} 
d_{t} \vec{W} +\nu _{e} \frac{\vec{\alpha }}{n_{e} } =
\left(\vec{W} \vec{\nabla } \right) \vec{V} ,
\end{equation} 
\[  \text{where }
\vec{W} \equiv \frac{\vec{\alpha }-\vec{\omega }_{He} }{n_{e} }.
\]
Note that in the case of a collisionless plasma ($\nu _{e}= 0$) equation (\ref{EQ14}) for the vector $\vec{W}$ formally coincides with the Helmholtz equation (see \cite{27,28}) for vorticity in the case of an incompressible fluid in the absence of a magnetic field $\vec {H}$: 
$d_{t} \vec{\alpha } = \left( \vec{\alpha} \vec{\nabla } \right) \vec{V} $.
Thus we have obtained a nonlinear vector equation describing the vortex dynamics of electrons.

Let us show that the vortical motion begins as soon as there appears a perturbation of the electron density. For transversal electron velocity $\vec{V}_{\bot } $~ from (\ref{EQ5}) one can obtain the following equation 
$$
\vec{V}_{\bot } =-\left(\frac{e}{m_{e} \omega _{He} } \right)\left[\vec{e}_{z} ,\; E_{or} \right]+
\left(\frac{e}{m_{e} \omega _{He} } \right)\left[\vec{e}_{z} ,\; \vec{\nabla }_{\bot } \phi \right]-
$$
\begin{equation}  
\label{EQ15} 
-\frac{1}{\omega _{He} } \partial _{t} \left[\vec{e}_{z} ,\; \vec{V}_{\bot } \right]-\frac{1}{\omega _{He} } \left[\vec{e}_{z} ,\; \left(\vec{V}_{\bot } \vec{\nabla }_{\bot } \right)\vec{V}_{\bot }\right],   
\end{equation} 
\[
\vec{\nabla}_{\bot} \varphi \equiv \vec{\nabla }_{\bot } \phi -E_{or},
\]
here - $\phi$ is the electric potential of the vortical perturbation, $E_{or}$ is the radial electrical field.

Taking into account higher linear terms, from equations (\ref{EQ15}) we derive 
 the following expression 
\[
\vec{V}_{\bot } \approx \vec{V}_{\theta o} +\left(\frac{e}{m_{e} \omega _{He} } \right)\left[\vec{e}_{z} ,\; \vec{\nabla }_{\bot } \varphi \right]+
\]
\begin{equation} 
\label{EQ16} 
+\left(\frac{e}{m_{e} \omega _{_{He} }^{2} } \right)\partial _{t} \vec{\nabla }_{\bot } \varphi.     
\end{equation} 
From (\ref{EQ3}), (\ref{EQ16}) we obtain the expression for the vorticity
\begin{equation} 
\label{EQ17} 
\alpha \approx \left(\frac{\omega _{_{pe} }^{2} }{\omega _{_{He} }^{} } \right)\left(\frac{\Delta n}{n_{oe} } \right)+\frac{e}{m_{e} } \partial _{t} \vec{e}_{z} \left[\vec{\nabla }_{\bot } ,\; \left(\frac{1}{\omega _{_{He} }^{2} } \right)\vec{\nabla }_{\bot } \varphi \right].       
\end{equation}

\noindent From (\ref{EQ17}) the expression for the unperturbed value of the vorticity follows
\begin{equation} 
\label{EQ18} 
\alpha _{o} \approx -\frac{2eE_{or} }{rm_{e} \omega _{He} } =\left(\frac{\omega _{_{pe} }^{2} }{\omega _{_{He} }^{} } \right)\left(\frac{\Delta n}{n_{oe} } \right).      
\end{equation} 

The expression (\ref{EQ17}) is convenient for physical interpretation of connection between the electron density perturbation $\delta n_{e} $ and the vortical motion. To see this, we consider the limiting case and neglect the ion motion. In this case from (\ref{EQ17}) it approximately follows
\begin{equation} 
\label{EQ19} 
\alpha \approx \left(\frac{\omega _{_{pe} }^{2} }{\omega _{_{He} }^{} } \right)\left(\frac{\Delta n}{n_{oe} } \right)+\left(\frac{\omega _{_{pe} }^{2} }{\omega _{_{He} }^{} } \right)\left(\frac{\delta n}{n_{oe} } \right).       
\end{equation} 

The first member in (\ref{EQ19}) specifies electron movement on the closed trajectories in the crossed fields. The second member in (\ref{EQ19}) shows, that the vortical motion begins as soon as there appears a perturbation of the electron density $\delta n_{e}$. From (\ref{EQ19}) it also follows, that vorticity has one sign in all beam area in the case of vortical perturbations of the small amplitudes. The opposite sign of the vorticity occurs in some regions of the beam, where the vortex amplitude exceeds a certain value.  

\section{Conclusions}

The Juno measurements of the aurora reveal valuable information about the precipitating particle population, which interact with the Jovian ionosphere plasma at varying altitudes. More detailed measurements of auroral structures from satellite trails in the infrared and ultraviolet ranges have been obtained \cite{29, 30}. Fine structure is observed on scale of approximately tens of kilometers. It reveal that in the case of the IFT trail, the emission has an alternating series of spots, reminiscent of vortices, at small distances from each other, and sometimes split into two arcs. The research of vortex structure formation mechanism can give an understanding of the reasons for these morphologies and of Jovian auroral processes.

In this paper the vortex dynamics of electron beams is considered in the framework of Io - Jupiter interaction. Currents flow from Io along the magnetic IFT and closed near the poles. These particles impact the ionosphere plasma and in the area of IFT footprint the auroras are generated.

The conditions of vortex perturbations formation, its properties and dynamics in the crossed configuration of the radial electric and longitudinal magnetic fields are described. The velocity of radial expansion of the electron beam due to electron collisions is taken into account. The nonlinear vector equation is obtained that describes the vortex dynamics of electrons. It is also analyzed how the vortex motion depends on the occurrence of electron density perturbations.

\vskip3mm 
 The  work was supported by Program of Fundamental Research of  Physics and Astronomy Department NAS of Ukraine (project  No.~0120U101347). The work was performed with the partial support of the Target Complex Program of the NAS of Ukraine on Plasma Physics.

\vspace*{-5mm} 
\rezume{
В.І.Маслов, О.К.Черемних, А.П.Фоміна, Р.І.Холодов, О.П.Новак, Р.Т.Овсянніков}
{Вихрові структури і динаміка електронного пучка в замагніченій плазмі} 
{В даній роботі ми досліджуємо задачу про формування вихрових структур при відбитті пучка електронів від подвійного шару іоносфери Юпітера. А також вплив цих структур на виникнення щільних висхідних електронних пучків, прискорених потенціалом подвійного шару вздовж токової трубки Іо, для яких стає можливий фазовий перехід в режим циклотронного надвипромінення. Розглянуто умови формування вихрових збурень. Знайдено нелінійне рівняння, яке описує вихрову динаміку електронів та вивчені його наслідки.}

\rezume{
В.И.Маслов, О.К.Черемных, А.П.Фомина, Р.И.Холодов, А.П.Новак, Р.Т.Овсянников}
{Вихревые структуры и динамика электронного пучка в замагниченной плазме} 
{В данной работе мы исследуем задачу о формировании  вихревых структур при отражении пучка электронов от двойного слоя ионосферы Юпитера. А также  влияние этих структур на возникновение плотных восходящих электронных пучков, ускоренных потенциалом двойного слоя вдоль токовой трубки Ио, для которых становится возможен фазовый переход в режим циклотронного сверхизлучения. Рассмотрены условия формирования вихревых возмущений. Найдено нелинейное уравнение, которое описывает вихревую динамику электронов  и изучены его следствия.}

\end{document}